\documentclass{article}

\usepackage{microtype}
\usepackage{graphicx}
\usepackage{subcaption}
\usepackage{booktabs} 

\usepackage{hyperref}

\usepackage[preprint]{icml2026}

\usepackage{amsmath}
\usepackage{amssymb}
\usepackage{mathtools}
\usepackage{amsthm}

\makeatletter

\expandafter\let\csname algorithm*\endcsname\relax
\expandafter\let\csname endalgorithm*\endcsname\relax
\makeatother
\usepackage[ruled,linesnumbered,vlined]{algorithm2e} 

\theoremstyle{plain}

\theoremstyle{definition}

\theoremstyle{remark}

\usepackage[textsize=tiny]{todonotes}
\usepackage{multirow}
\icmltitlerunning{Submission and Formatting Instructions for ICML 2026}

\begin{document}
\abovedisplayskip=8pt plus 2pt minus 4pt
\belowdisplayskip=8pt plus 2pt minus 4pt
\abovedisplayshortskip=4pt plus 2pt
\belowdisplayshortskip=4pt plus 2pt minus 2pt
\newcommand{\system}{$\textsc{OptiLeak}$\xspace}
\newcommand{\daux}{$\mathcal{D}_{\text{aux}}$}
\twocolumn[

  \icmltitle{\system: Efficient Prompt Reconstruction via Reinforcement Learning in Multi-tenant LLM Services}



  \icmlsetsymbol{equal}{*}

  \begin{icmlauthorlist}
    \icmlauthor{Longxiang Wang}{1}
    \icmlauthor{Xiang Zheng}{1}
    \icmlauthor{Xuhao Zhang}{2}
    \icmlauthor{Yao Zhang}{2}
    \icmlauthor{Ye Wu}{2}
    \icmlauthor{Cong Wang}{1}
  \end{icmlauthorlist}

  \icmlaffiliation{1}{City University of Hong Kong}
  \icmlaffiliation{2}{ByteDance Inc.}
  \icmlcorrespondingauthor{Cong Wang}{congwang@cityu.edu.hk}
  \icmlcorrespondingauthor{Xiang Zheng}{xiang.zheng@cityu.edu.hk}

  \icmlkeywords{Machine Learning, ICML}

  \vskip 0.3in
]


%

\newcommand{\fix}{\marginpar{FIX}}
\newcommand{\new}{\marginpar{NEW}}

\newcommand{\xz}[1]{\refstepcounter{note}{\bf\textcolor{orange}{$\ll$Xiang~\thenote: {\sf #1}$\gg$}}}
\newcommand{\rev}[1]{\textcolor{black}{#1}}

\printAffiliationsAndNotice{}


\begin{abstract}

Multi-tenant LLM serving frameworks widely adopt shared Key-Value caches to enhance efficiency. However, this creates side-channel vulnerabilities enabling prompt leakage attacks. Prior studies identified these attack surfaces yet focused on expanding attack vectors rather than optimizing attack performance, reporting impractically high attack costs that underestimate the true privacy risk.
We propose \system, a reinforcement learning-enhanced framework that maximizes prompt reconstruction efficiency through two-stage fine-tuning. 
Our key insight is that domain-specific ``hard tokens''---terms difficult to predict yet carrying sensitive information---can be automatically identified via likelihood ranking and used to construct preference pairs for Direct Preference Optimization, eliminating manual annotation.
This enables effective preference alignment while avoiding the overfitting issues of extended supervised fine-tuning.
Evaluated on three benchmarks spanning medical and financial domains, \system achieves up to $12.48\times$ reduction in average requests per token compared to baseline approaches, with consistent improvements across model scales from 3B to 14B parameters.
Our findings demonstrate that cache-based prompt leakage poses a more severe threat than previously reported, underscoring the need for robust cache isolation in production deployments.

\end{abstract}
\section{Introduction}
Large Language Models (LLMs) have been widely deployed across various domains, from Chatbots~\cite{openai-chatbot,claude-chatbot} to GUI agents~\cite{ms-guiagent}. To improve inference efficiency and reduce computational costs, modern LLM service frameworks (e.g., vLLM~\cite{vllm-kwon-23}, SGLang~\cite{sglang-zheng-23}) share Key-Value (KV) cache entries across user requests.
Recently, both academia and industry have begun exploring persistent KV-cache storage to enhance cache reusability and reduce memory costs. For instance, \cite{infinigen-lee-24,shadowkv-sun-24} propose offloading KV-cache from GPU to CPU memory to alleviate expensive GPU memory usage. Similarly, vLLM integrates such KV-cache management approaches into its framework\cite{vllm-cache-storage}. As storage capacity becomes increasingly affordable and cache persistence more prevalent, it becomes urgent to understand and manage the privacy risks of cached queries.

Prior research has shown that cache-sharing strategies in LLM services can introduce different types of side channels, such as Time to First Token (TTFT)~\cite{inputsnatch-zheng-24,earlybird-song-24} and ordering based on LLM scheduling policies~\cite{iknowwhat-wu-25}.
These side channels enable adversaries to craft malicious requests for revealing matches with other users and recover sensitive user query information. 
\rev{However, these studies primarily focused on exploring the expanded attack surface within cache sharing mechanisms, overlooking the practical efficiency of side-channel exploitation.
Prior works~\cite{auditing-gu-25,shadowinthecache-luo-25} indicate that extracting information typically requires an impractically large number of requests.
This inefficiency stems from not fully accounting for a real-world adversary's capability to optimize the attack strategy.
Consequently, the true severity of these privacy risks remains underestimated.}
This gap hinders accurate assessment of leakage severity and the development of countermeasures.

To fill this gap, we investigate how to maximize real adversary attack capabilities through model optimization in this paper.
We consider a practical threat model where the adversary is assumed to have knowledge of the general domain of user queries.
This information can be inferred through various means, such as application context, service endpoints, or metadata.
This assumption about the adversary's knowledge is realistic, especially given the increasing number of organizations that deploy private LLMs or utilize private LLM inference services from third-party cloud providers~\cite{financial-example,law-example,llmhealthcare-Dennstadt-25}.
Users of these domain-specific LLM services typically inquire about topics within specific domains, e.g., financial analysis, medical diagnosis, or legal research.
Consequently, attackers can leverage public datasets from similar domains to finetune their local models, allowing for efficient side-channel attacks.

Training in a domain-aware setting presents two main challenges.  First, Supervised Fine-Tuning (SFT) tends to overfit and struggles to effectively learn domain-specific knowledge since specific-domain QA datasets are still dominated by general language, which makes it difficult for the model to grasp specialized terminology.
Second, Reinforcement Learning (RL) algorithms, e.g., Proximal Policy Optimization (PPO)~\cite{ppo-schulman-17} and Group Relative Policy Optimization (GRPO)~\cite{grpo-shao-24}, face the complexity of reward engineering. 
Rewards based on semantic similarity can lead to overfitting on superficial patterns, while rewards that rely on exact token matching tend to provide insufficient feedback.

To enable preference alignment for domain-specific knowledge while overcoming the challenging reward design, we propose \system, a two-stage fine-tuning framework with a novel automated annotation approach.
In the first stage, we apply SFT on the base local model using domain-specific data to familiarize it with relevant knowledge in that specific field.
In the second stage, we leverage this SFT-tuned model to automatically annotate the training data for DPO.
This automatic annotation involves identifying what we refer to as ``hard tokens'', which are domain-specific terms that are challenging to generate but contain crucial and sensitive domain information.
In this way, \system eliminates the need for manual annotation and 
prioritizes the extraction of genuinely sensitive, domain-specific information, guided by feedback on the identified hard tokens.

Through evaluation, we show that \system achieves significant performance improvements across three domain-specific datasets: MedQA~\cite{medqa-di-21}, PubMedQA~\cite{pubmedqa-jin-19}, and FinanceBench~\cite{financebench-islam-23}.
Compared to the baseline approach that uses base models as local LLMs for users' prompt reconstruction~\cite{iknowwhat-wu-25}, \system achieves a 12.48$\times$ reduction in Average Requests Per Token (ARPT) on FinanceBench with Qwen2.5-3B-Instruct~\cite{qwen2.5-yang-24} as the backbone and also demonstrates substantial performance gains on MedQA and PubMedQA.
Furthermore, we investigate the impact of data distribution similarity on attack efficiency. Finally, we present an ablation study analyzing the relationship between SFT overfitting and adversarial performance, further validating the effectiveness of \system.

In summary, our contributions are as follows:
\begin{itemize}
\item We formalize adversary optimization for cache-based prompt leakage attacks and demonstrate that optimized attackers pose significantly greater risk than previously reported, achieving up to $12.48\times$ improvement in attack efficiency.
\item We propose \system, a two-stage framework combining SFT with DPO, featuring an automated annotation mechanism that identifies hard tokens for preference learning without manual labeling.
\item We conduct comprehensive evaluations on three benchmarks across two knowledge-intensive domains (medical and finance) to demonstrate \system's effectiveness and provide quantitative analysis of performance improvements.
\item We provide extensive discussion of \system's compatibility with other active side channels, and potential application as a proactive side channel risk assessment tool for LLM service providers.
\end{itemize}

\section{Related Works}

\textbf{Side Channel Attacks in LLM Services.}
Current side channel attacks in LLM services can be divided into two types: passive side channel attacks~\citep{whatwasyour-weiss-24,timewilltell-zhang-24,privacyrisk-wei-24} and active side channel attacks~\citep{iknowwhatyousaid-gao-25,earlybird-song-24,iknowwhat-wu-25,inputsnatch-zheng-24,spillthebeans-adiletta-25,shadowinthecache-luo-25,invalidatecompare-zhang-24}.
While passive attacks focus on monitoring encrypted traffic or timing patterns (detailed in Appendix~\ref{app:extended_related_works}), our work focuses on active side channel attacks.

In an active side channel attack, the attacker actively interacts with or manipulates the victim LLM or its underlying system (e.g., hardware caches or serving mechanisms) during the user's query process.
For example, ~\citet{spillthebeans-adiletta-25} and ~\citet{iknowwhatyousaid-gao-25} demonstrate that attackers can infer user queries by monitoring hardware cache changes to determine the position of accessed embedding vectors.
\citet{inputsnatch-zheng-24} leverages timing-based side channels by detecting whether queries hit the cache to determine if proposed dummy queries match users' actual queries.
\citet{iknowwhat-wu-25} demonstrates that serving ordering mechanisms (e.g., longest prefix matching) also reveal side channel information when proposed queries match with targeted ones.
Our work distinguishes itself from previous works by employing RL to enhance cache-based active side channel attacks.
Rather than expanding attack scenarios, we focus on optimizing the probing query generation process to explore the maximum capacities of real-world adversaries.

\textbf{RL-Based Red-Teaming for LLMs.}
RL-based red teaming has emerged as a prominent paradigm for discovering vulnerabilities of LLMs in security and privacy.
A series of works~\cite{advprompter-paulus-24,rljack-chen-24,swordecho-tang-24} leverage RL for enhancing jailbreaking techniques, enabling effective evasion of the victim LLM's guardrail and alignment to generate malicious content.
Similarly, RL is employed to prompt LLMs to produce harmful responses that incorporate toxic content, sensitive terminology, or factual errors~\cite{redteaming-perez-22,exploreestablish-casper-23,calm-zheng-25,diverct-zhao-25}. 
Recently, RL is applied to privacy leakage attacks against LLMs.
\citet{leakagent-chen-25} utilizes RL for system prompt extraction~\cite{extractingtraining-carlini-21,llmcanbe-li-22,decodingtrust-wang-23} and training data extraction attacks~\cite{scalableextraction-nasr-25} with word edit similarity as a reward function to elicit private information from the victim LLM.
In contrast to previous research, \system target a distinct privacy attack scenario,  leveraging RL to optimize adversarial strategies for recovering users' real-time prompts through cache-sharing side channels.

\section{Problem Formulation}
In this section, we describe the attacking surface of the LLM serving system (i.e., side channel by cache sharing), our threat model, and the detailed attack scenario.

\subsection{Side Channel by Cache Sharing Mechanism}

Current multi-tenant LLM service frameworks (e.g., vLLM~\cite{vllm-kwon-23}, SGLang~\cite{sglang-zheng-23}, LightLLM~\cite{tensorcompiler-nakandala-20}) are primarily based on decoder-only Transformer~\cite{attentionisall-vaswani-17} architecture. 
During autoregressive inference where LLMs generate text token-by-token, the KV cache mechanism stores computed key and value vectors for previously processed tokens, eliminating redundant calculations in subsequent decoding steps.
Due to the causal nature of the attention mask in LLMs, the key and value representations for a given token depend only on the preceding tokens in the sequence. This property enables KV cache sharing across different requests: when multiple users submit prompts with identical prefixes, the cached key-value pairs for these shared prefix tokens can be reused, significantly reducing both memory consumption and computational overhead~\cite{vllm-kwon-23,sglang-zheng-23}.
The cache sharing architecture operates through a unified memory pool where requests from different tenants can access shared cached states. When a cache hit occurs, the system directly utilizes stored values, thereby reducing TTFT and overall response latency. However, while this optimization accelerates inference performance, it concurrently introduces security vulnerabilities. The shared nature of these caches creates an observable timing side-channel, as cache hits result in substantially faster response times compared to cache misses, potentially leaking information about other users' inputs through timing analysis.

\subsection{Threat Model}
\label{subsec: threat model}

In our threat model, the adversary's goal is to reconstruct prompts sent by other victim clients through cache-based side channels. Following the assumptions established by~\citet{iknowwhat-wu-25} and~\citet{inputsnatch-zheng-24}, the adversary possesses capabilities equivalent to an ordinary LLM client, with only black-box access to the LLM server (i.e., the adversary has no privilege to access the model architecture or parameters). The LLM operates in streaming mode, delivering responses to users in real time. 
We assume the adversary can: (1) send queries (with a maximum limit for each victim query) to the LLM server, and correspondingly measure TTFT to detect cache hits; (2) access publicly available tokenizers, which are commonly provided by online LLM services such as~\citet{openai-tokenizer} and open-source LLMs~\citep{qwen-tokenizer}. The adversary cannot view internal server logs, modify server configurations, or access other users' communication channels directly.

We further assume the adversary has domain knowledge about victim clients' queries (e.g., awareness that a user is a doctor who probably submits medical-related queries) but lacks knowledge on specific prompt templates or the exact query content used by the victim client~\cite{iknowwhat-wu-25,wiretappingllm-soleimani-25}. This assumption is practical, as more and more data-sensitive institutions (e.g., Medical~\cite{llmhealthcare-Dennstadt-25}, Legal~\cite{law-example}, Financial~\cite{financial-example}) choose to privately deploy LLMs or lease private LLM services from third-party cloud providers. For example, in a hospital's LLM deployment, an adversary (such as a malicious staff member or a compromised account) can assume that most queries are medical-related, but still cannot access the specific content of other users' sensitive patient information.

\begin{figure}
    \centering
    \includegraphics[width=0.95\linewidth]{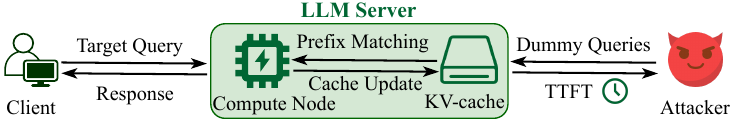}
   \caption{Attack scenario overview}
    \label{fig:attack-scenario}
\end{figure}

\subsection{Attack Scenario}
Figure~\ref{fig:attack-scenario} illustrates our attack scenario in the context of cache-based prompt leakage attacks. Our system consists of two parties: victim clients and a cloud-based LLM inference server. The server concurrently receives queries from multiple users, with these requests being processed on shared GPU nodes. 
\rev{We assume that the victim's and attacker's queries are processed on a server node sharing the same KV cache pool. This is a realistic assumption, as many production deployments (e.g., Google~\cite{google-kvcache}, Microsoft~\cite{adaptcache-feng-25}, ByteDance~\cite{bytedance-kvcache}) currently employ KV cache pooling to persist cached key-value pairs over extended periods. Compared to traditional physical GPU node allocation, the pooling architecture increases the likelihood that attackers and victims share the same KV cache node. Moreover, this assumption aligns with prior work on KV cache side-channel attacks~\cite{iknowwhat-wu-25,inputsnatch-zheng-24,earlybird-song-24}.}

At a high level, the attack process operates as follows: (1) The victim sends the target query to the LLM server, and the query's tokens are stored in the KV-cache. (2) The adversary then sends multiple specifically-designed dummy queries to the LLM server, attempting to match prefixes of the target query. (3) The adversary observes each response's TTFT to determine whether a cache hit occurs, which indicates whether the queried tokens match the targeted ones. \system operates iteratively, with each iteration recovering one token from the victim's prompt. The adversary terminates when either the entire targeted prompt is reconstructed or the maximum number of attempts is reached.

\section{Methodology}

\begin{figure*}
    \centering
    \includegraphics[width=0.91\linewidth]{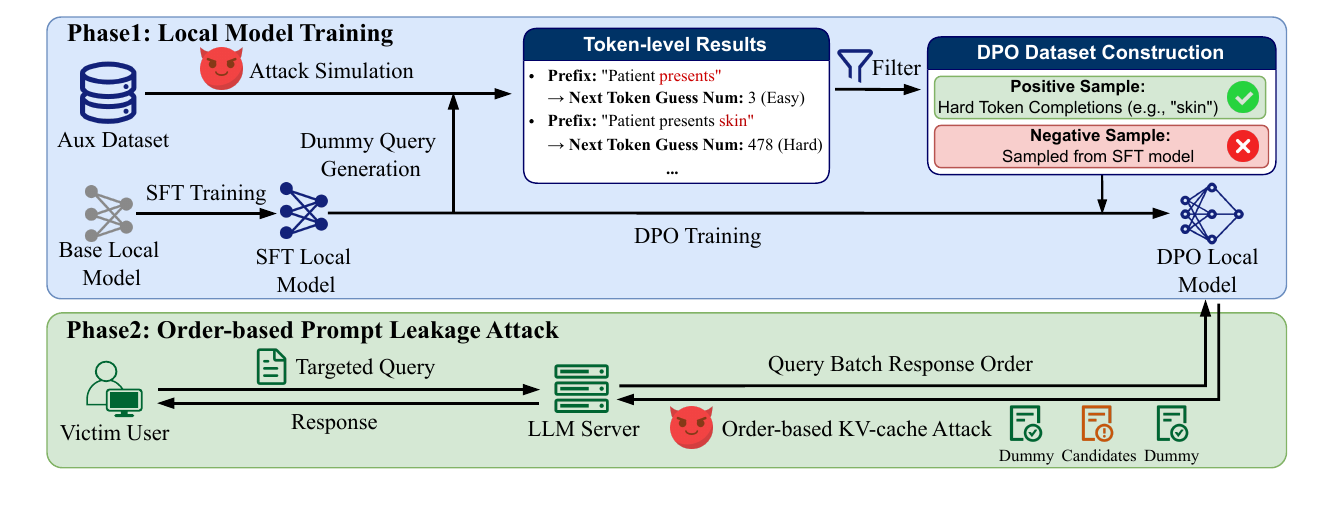}
    \caption{An overview of \system's operation pipeline.}
    \label{fig: system overview}
\end{figure*}

We now introduce \system, an automated prompt leakage attack framework that leverages Supervised Fine-Tuning (SFT) and Direct Preference Optimization (DPO) to fine-tune an adversarial model for probe query generation. We first describe our RL-based approach for automatically optimizing adversary capabilities and then present the order-based KV-cache side channel attack. 
\rev{Figure~\ref{fig: system overview} provides an overview of \system. The corresponding operational steps are detailed in Algorithm~\ref{alg:Operation pipeline} within Appendix~\ref{appendix: algorithm procedure}.}

\subsection{Local Model Fine-Tuning}
\label{local model finetune}
\system involves a two-stage approach for local model fine-tuning in \system, i.e., first SFT then DPO. We assume an auxiliary dataset for SFT based on the adversary's domain knowledge, as mentioned in Section \ref{subsec: threat model}. For DPO, we propose a novel automated annotation mechanism.

\paragraph{Supervised Fine-Tuning.} Given an auxiliary dataset $\mathcal{D}_\text{aux} = \{( x,y)\}$, where $x$ and $y$ are user query and LLM response separately, we first apply SFT to the base adversarial model $\pi_{\text{base}}$ via maximizing the likelihood of the user query and the corresponding response:

\begin{equation}
\mathcal{L}_{\text{SFT}}(\theta) = \sum_{s \in \mathcal{D}_{\text{aux}}} \sum_{t=1}^{|s|} \log \pi_{\theta}(s_t | s_{<t}),
\end{equation}
where $s=[x,y]=[s_{<t},s_t]$ denotes the concatenated sequence of user query and LLM response, $|s|$ is the number of tokens in $s$.

\paragraph{Direct Preference Optimization.}
After SFT, we obtain $\pi_{\text{SFT}}$. Though extending training epochs of SFT is a traditional approach for performance improvement, we find in our experiment that increasing SFT training epochs leads to poor attack performance, primarily due to overfitting and mode collapse.
\rev{As demonstrated in our ablation study (Figure~\ref{fig:ablation_study_results}), extending SFT training causes the number of guessing attempts to increase to $3.70\times$ compared with the baseline.
To further improve the attacker's capacity, we leverage DPO for better alignment of the adversarial model with the domain-specific dataset.}
Our key insight stems from the observation that there is a small subset of queries that are challenging to be predicted.
This motivates us to develop an automated mechanism to identify these challenging queries for 
better preference alignment, thereby enhancing overall model capacity.

\paragraph{Automated Annotation for DPO.} The \emph{automated annotation} approach we propose for DPO leverages token prediction difficulty (i.e., likelihood of each token) as a signal for constructing preference pairs.
Our approach begins by identifying hard-to-recover tokens. Specifically, we use $\pi_{\text{SFT}}$ to predict each token $s_t$ in each sample $s = [x,y]$ of the auxiliary dataset $\mathcal{D}_{\text{aux}}$ and rank the target token in the descending order based on its likelihood $\pi_\text{SFT}(s_t|s_{<t})$.
Intuitively, this ranking represents the recovery difficulty, where tokens with higher rankings indicate higher difficulty.
Formally, we identify all preferred responses $s^\text{win}$ via
\begin{equation}
\{s^\text{win}=[s^\text{win}_{<t}, s^\text{win}_t]\,|\,[s_{<t}, s_t]\,\forall s \in \mathcal{D}_\text{aux} \,\text{if}\, \mathrm{Rank}(s_t) > \gamma\},
\end{equation}
where $\mathrm{Rank}(s_t)$ is a function that returns the likelihood ranking of $s_t$ among the likelihood set $\{\pi_\text{SFT}(v|s_{<t})|v\in\mathcal{V}\}$ ($\mathcal{V}$ is the vocabulary).
Then, for each prefered resposne $s^\text{win}=[s^\text{win}_{<t}, s^\text{win}_t]$, we construct the corresponding dispreferred response $s^\text{lose}$  using $\pi_{\text{SFT}}$ via greedy decoding at the position of the hard token $s_t$:
\begin{equation}
s^\text{lose} = [s^\text{lose}_{<t}, s^\text{lose}_{t}] = [s^\text{win}_{<t}, \arg\max_{v \in \mathcal{V}} \pi_{\text{SFT}}(v | s^\text{win}_{<t})],
\end{equation}
That is, $s^\text{lose}_{<t}=s^\text{win}_{<t}$ and $s^\text{lose}_{t}=\arg\max_{v \in \mathcal{V}} \pi_{\text{SFT}}(v | s^\text{win}_{<t})$ (the greedy decoding operator). Through this, we create pairs $(s^\text{win}, s^\text{lose})$ where each $s^\text{win}$ contains a hard token $s^\text{win}_t$, while the corresponding $s^\text{lose}$ contains a highly confident yet incorrect token at the same position.
To clarify this, we provide an example in Figure~\ref{fig: system overview}. Given a samlpe $s$ ``Patient presents skin rash'' in the dataset, we identify ``skin'' is a hard token since its likelihood ranking is the 478-th (exceeding our threshold) and thus form a preference pair: prefix  ``Patient presents'' as $s^\text{win}_{<2}$, the preferred token ``skin'' as $s^\text{win}_2$, and the dispreferred token $s^\text{lose}_2$ sampled from SFT-tuned model.

We then construct the preference dataset $\mathcal{D}_{\text{pref}} = \{(s^\text{win},s^\text{lose})\}$
and apply DPO~\cite{dpo-rafailov-23} using $\mathcal{D}_{\text{pref}}$ to improve $\pi_{\text{SFT}}$. The loss function for DPO is:
\begin{equation}
{\small
\begin{split}
\mathcal{L}_{\text{DPO}}(\theta) = -\mathbb{E}_{(s^\text{win},s^\text{lose}) \sim \mathcal{D}_{\text{pref}}} \biggl[ \log \sigma \Bigl( \beta \log \frac{\pi_\theta(s^\text{win}_{t}|s^\text{win}_{<t})}{\pi_{\text{ref}}(s^\text{win}_{t}|s^\text{win}_{<t})} \\
- \beta \log \frac{\pi_\theta(s^\text{lose}_{t}|s^\text{lose}_{<t})}{\pi_{\text{ref}}(s^\text{lose}_{t}|s^\text{lose}_{<t})} \Bigr) \biggr],
\end{split}
}
\end{equation}
where $\pi_{\text{ref}}$ is the reference model in DPO that is initialized from $\pi_{\text{SFT}}$ and remains fixed during training,
$\beta$ is a hyperparameter that controls the strength of the KL regularization, and $\sigma$ is the sigmoid function. This approach maximizes the likelihood of each preferred response $s^\text{win}$ relative to its corresponding dispreferred responses $s^\text{lose}$ while maintaining the model's overall capabilities through KL regularization with respect to $\pi_{\text{ref}}$.

\subsection{Order-based KV-cache Prompt Leakage Attack}
Let the victim user's targeted query we want to recover be $s^\text{victim}$, which contains $n$ tokens.
To launch a specific prompt leakage attack, we leverage the Longest Prefix Match (LPM) scheduling policy utilized by SGLang~\cite{sglang-zheng-23}, one of the most prominent LLM inference frameworks. This policy ensures that when the request queue contains multiple queries, waiting requests are prioritized based on the length of their matched prefix tokens. 
To guess a token $s^\text{victim}_t$, we send a batch of queries $Q = \{\tilde{q}^1,...,\tilde{q}^k\}$ to the LLM server, where $k = 2m + |Q_\text{gen}|$ represents the total number of queries. 
The batch contains $Q_\text{gen}$ candidate queries (our generated guesses for position $i$) placed between two groups of $m$ dummy queries each. Each dummy query shares the same low-probability token obtained from local LLM predictions.
All generated queries share the same prefix $q_{<t}$ and differ only in the last token $q_t$. After sending the entire batch, we observe the response order with TTFT to determine if a cache hit occurs.
Specifically, if a cache hit occurs, the query containing the correct token match will be prioritized due to LPM scheduling, causing a gap between consecutive real queries in the response sequence. To detect cache hit $\mathsf{D_{hit}}$, we use:
\begin{equation}
{\small
 \mathsf{D_{hit}} = \begin{cases}
1, & \exists \tilde{q}^i \in Q_{\text{gen}}\ \text{s.t.}\
\begin{aligned}[t]
&\left[\min_{\tilde{q} \in Q_{\text{gen}} \setminus \{\tilde{q}^i\}} \text{pos}(\tilde{q})\right] \\
&- \text{pos}(\tilde{q}^i) > \lceil \theta \cdot m \rceil
\end{aligned} \\
0, & \text{otherwise},
\end{cases}
}
\end{equation}
where $\text{pos}(\tilde{q})$ is the response position of query $\tilde{q}$, $\theta \in (0,1)$ is a parameter to adjust the sensitivity of cache hit detection. When $\mathsf{D_{hit}} = 1$, we identify $\tilde{q}^i$ contains the correct token.
\section{Experiment}

\subsection{Experimental Setup} 
\rev{We implement and evaluate \system on a Linux-based operating system, and more detailed experiment settings are described below. For our implementation, we leverage SGLang~\cite{sglang-zheng-23} as the LLM serving and inference framework.}

\paragraph{Baselines \& Benchmarks.}
We conduct evaluation using four LLMs from two series: Qwen-2.5 (3B, 7B, 14B)\cite{qwen2.5-yang-24} and Llama-3.1-8B\cite{llama3-llama-24}. We use three real-world domain-specific datasets in the evaluation: MedQA~\cite{medqa-di-21}, PubMedQA~\cite{pubmedqa-jin-19}, and FinanceBench~\cite{financebench-islam-23}. Detailed descriptions of the baselines and benchmarks are provided in Appendix~\ref{appendix: baseline setup}.

\begin{table*}[t]
\centering
\caption{\rev{Main experimental results on MedQA, FinanceBench, and PubMedQA datasets. DPO results are obtained by directly training the base model using \system's auto-annotation approach. The best-performing results in the table are highlighted for clarity.}}
\resizebox{\textwidth}{!}{%
\begin{tabular}{@{}cc|ccccc|ccccc|ccccc@{}}
\toprule
\multirow{2}{*}{Model} & \multirow{2}{*}{Method} & \multicolumn{5}{c|}{MedQA} & \multicolumn{5}{c|}{FinanceBench} & \multicolumn{5}{c}{PubMedQA} \\
\cmidrule(lr){3-7} \cmidrule(lr){8-12} \cmidrule(lr){13-17}
& & ASR$_{500}$  & ASR$_{1000}$ & ASR$_{10000}$ & w/l & ARPT & ASR$_{500}$ & ASR$_{1000}$ & ASR$_{10000}$ & w/l & ARPT & ASR$_{500}$ & ASR$_{1000}$ & ASR$_{10000}$ & w/l & ARPT \\
\midrule
\multirow{4}{*}{Qwen-2.5-3B-Instruct} & Base & 18.0\%& 37.3\%& 95.3\%& - & 41.99 & 2.0\%& 2.0\%& 54.0\%& - & 240.81 & 0.0\% & 3.0\% & 68.0\% & - & 628.50 \\
& SFT & 27.3\%& 53.3\%& 98.0\%& 80.0\% & 26.37 &\textbf{88.0\%}&\textbf{90.0\%}&\textbf{100.0\%}& \textbf{100.0\%} & 20.81 & \textbf{12.0\%} & 28.0\% & \textbf{94.0\%} & \textbf{91.0\%} & 220.85 \\
& DPO & 18.0\%& 38.0\%& 95.3\%& 48.7\% & 41.86 & 2.0\%& 2.0\%& 56.0\%& 68.0\% & 238.81 & 1.0\% & 5.0\% & 71.0\% & 63.0\% & 571.51 \\
\cmidrule(lr){2-7} \cmidrule(lr){8-12} \cmidrule(lr){13-17}
& \system & \textbf{29.3\%}& \textbf{55.3\%}& \textbf{98.6\%}& \textbf{84.0\%} & \textbf{21.60} & \textbf{88.0\%}& \textbf{90.0\%}& \textbf{100.0\%}& \textbf{100.0\%} & \textbf{19.29} & 11.0\% & \textbf{31.0\%} & \textbf{94.0\%} & \textbf{91.0\%} & \textbf{208.72} \\
\midrule
\multirow{4}{*}{Qwen-2.5-7B-Instruct} & Base & 18.0\%& 36.0\%& 94.0\%& - & 54.50 & 6.0\%& 6.0\%& 66.0\%& - & 259.65 & 0.0\% & 1.0\% & 49.0\% & - & 943.05 \\
& SFT & \textbf{24.0\%}& \textbf{49.3\%}& 94.7\%& 79.3\% & 44.50 & 76.0\%& 82.0\%& \textbf{96.0\%}& 94.0\% & 69.63 &\textbf{9.0\%} & 21.0\% & 86.0\% & 91.0\% & 291.28 \\
& DPO &17.3\% &37.3\% & 94.7\%& 55.3\% & 52.67 & 6.0\%& 6.0\%& 72.0\%& 84.0\% & 246.38 & 0.0\% & 1.0\% & 49.0\% & 70.0\% & 901.09 \\
\cmidrule(lr){2-7} \cmidrule(lr){8-12} \cmidrule(lr){13-17}
& \system & 22.7\%& 48.0\%& \textbf{96.0\%}& \textbf{81.3\%} & \textbf{40.49} &\textbf{86.0\%}& \textbf{92.0\%}& \textbf{96.0\%}& \textbf{96.0\%} & \textbf{66.87} & 6.0\% & \textbf{22.0\%} & \textbf{88.0\%} & \textbf{91.0\%} & \textbf{274.91} \\
\midrule
\multirow{4}{*}{Qwen-2.5-14B-Instruct} & Base &23.3\% &45.3\% & 94.7\%& - & 44.35 &4.0\% &4.0\% & 64.0\%& - & 214.19 & 0.0\% & 0.0\% & 58.0\% & - & 796.08 \\
& SFT &\textbf{31.3\%} &56.6\% & 96.0\%& 78.6\% & 33.69 & 78.0\%& \textbf{88.0\%}& \textbf{98.0\%}& 96.0\% & 39.49 & 4.0\% & 24.0\% & 81.0\% & 88.0\% & 374.85 \\
& DPO & 24.0\%&46.0\% & 94.7\% & 68.0\% & 43.83 & 4.0\%& 4.0\%& 64.0\%& 62.0\% & 213.46 & 0.0\% & 0.0\% & 59.0\% & 40.0\% & 753.28 \\
\cmidrule(lr){2-7} \cmidrule(lr){8-12} \cmidrule(lr){13-17}
& \system & 30.7\% & \textbf{60.0\%}& \textbf{97.3\%}& \textbf{82.0\%} & \textbf{30.60} & \textbf{84.0\%}& \textbf{88.0\%}& \textbf{98.0\%}& \textbf{98.0\%} & \textbf{36.31} & \textbf{12.0\%} & \textbf{30.0\%} & \textbf{89.0\%} & \textbf{88.0\%} & \textbf{304.76} \\
\midrule
\multirow{4}{*}{Llama-3.1-8B-Instruct} & Base & 5.3\%& 27.3\%& 92.0 \%& - & 60.00 & 0.0\%& 0.0\%& 60.0\%& - & 286.52 & 0.0\% & 0.0\% & 57.0\% & - & 801.71 \\
& SFT & 8.7\%& 43.3\%& 98.0\%& 74.0\% & 30.43 & 62.0\%& \textbf{86.0\%}& 98.0\%& \textbf{98.0\%} & 37.91 &  13.0\% & \textbf{33.0\%} & \textbf{93.0\%} & \textbf{94.0\%} & 254.86 \\
& DPO & 8.0\%& 28.7\%& 92.7\%& 74.0\% & 54.86 & 0.0\%& 2.0\%& 66.0\%& 88.0\% & 248.69 & 0.0\% & 0.0\% & 60.0\% & 84.0\% & 666.01 \\
\cmidrule(lr){2-7} \cmidrule(lr){8-12} \cmidrule(lr){13-17}
& \system & \textbf{12.6\%}& \textbf{48.7\%}& \textbf{98.7\%}& \textbf{79.3\%} & \textbf{28.31} & \textbf{64.0\%}& \textbf{86.0\%}& \textbf{100.0\%}& \textbf{98.0\%} & \textbf{33.89} & \textbf{14.0\%} & \textbf{33.0\%} & \textbf{93.0\%} & \textbf{94.0\%} & \textbf{241.48} \\
\bottomrule
\end{tabular}%
}
\label{tab:main results}
\end{table*}

\paragraph{Evaluation Metrics.}
In the evaluation, we specifically employ three metrics:
\begin{itemize}
    \item \textit{Attack Success Rate} (ASR)  measures the success rate of prompt leakage attacks. To more precisely represent the adversary's capacity and account for the characteristics of real-world KV-cache systems, we adopt different maximum request limits. 
    For instance, ASR$_{10000}$ denotes a limit of 1000 request attempts.
    \item \textit{Average Requests Per Token} (ARPT) tracks the average number of effective requests required to recover a single token. This metric provides an overall assessment of the impact of model optimization methods at the token level.
    \item \textit{Request Number Win/Lose Rate} (w/l) measures the proportion of queries where the optimized model requires fewer requests compared to the base model. 
\end{itemize}

\begin{figure*}[t]
    \centering
    \begin{subfigure}{0.24\linewidth}
        \centering
        \includegraphics[width=\linewidth]{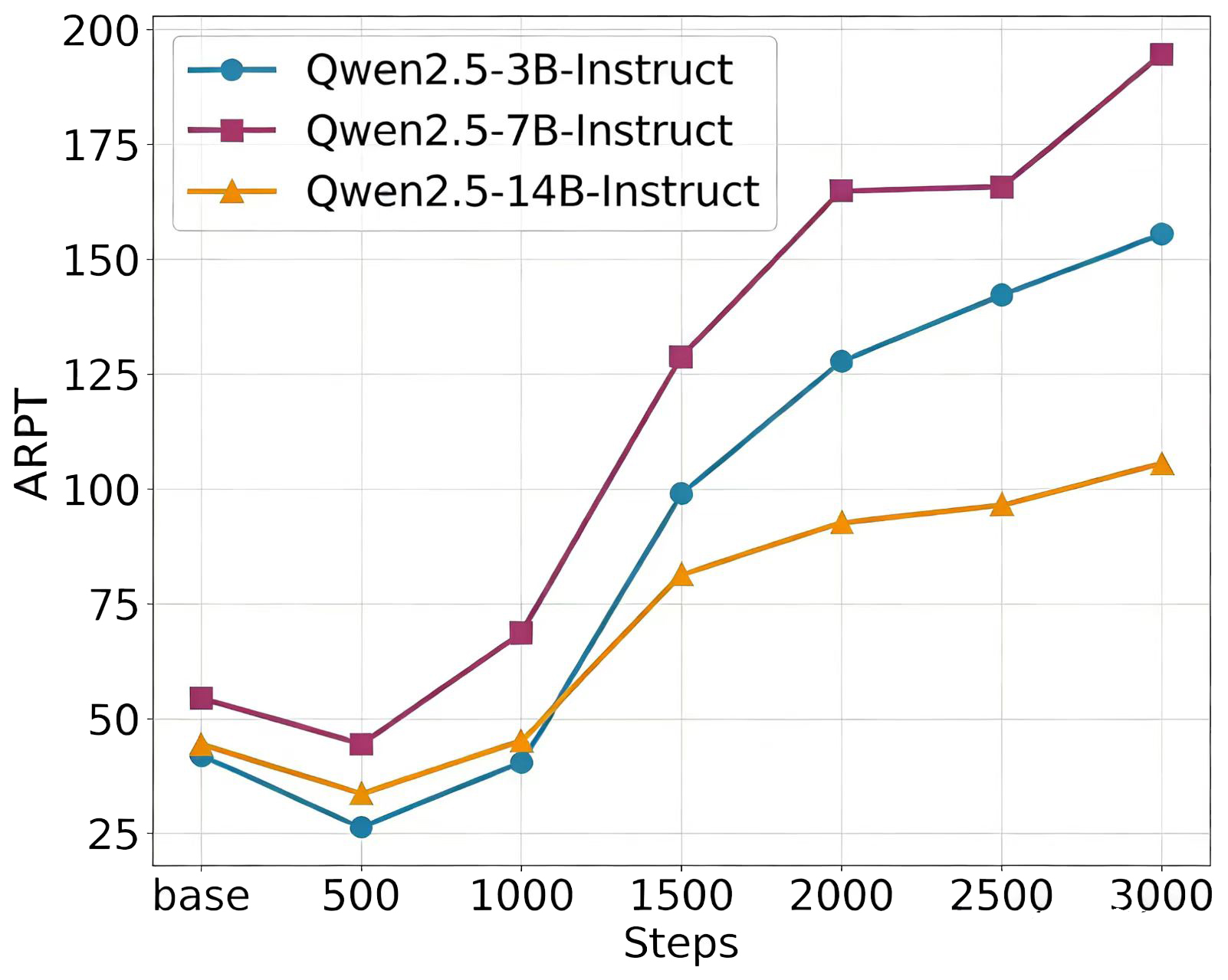}
        \caption{MedQA, SFT}
    \end{subfigure}
    \hfill
    \begin{subfigure}{0.24\linewidth}
        \centering
        \includegraphics[width=\linewidth]{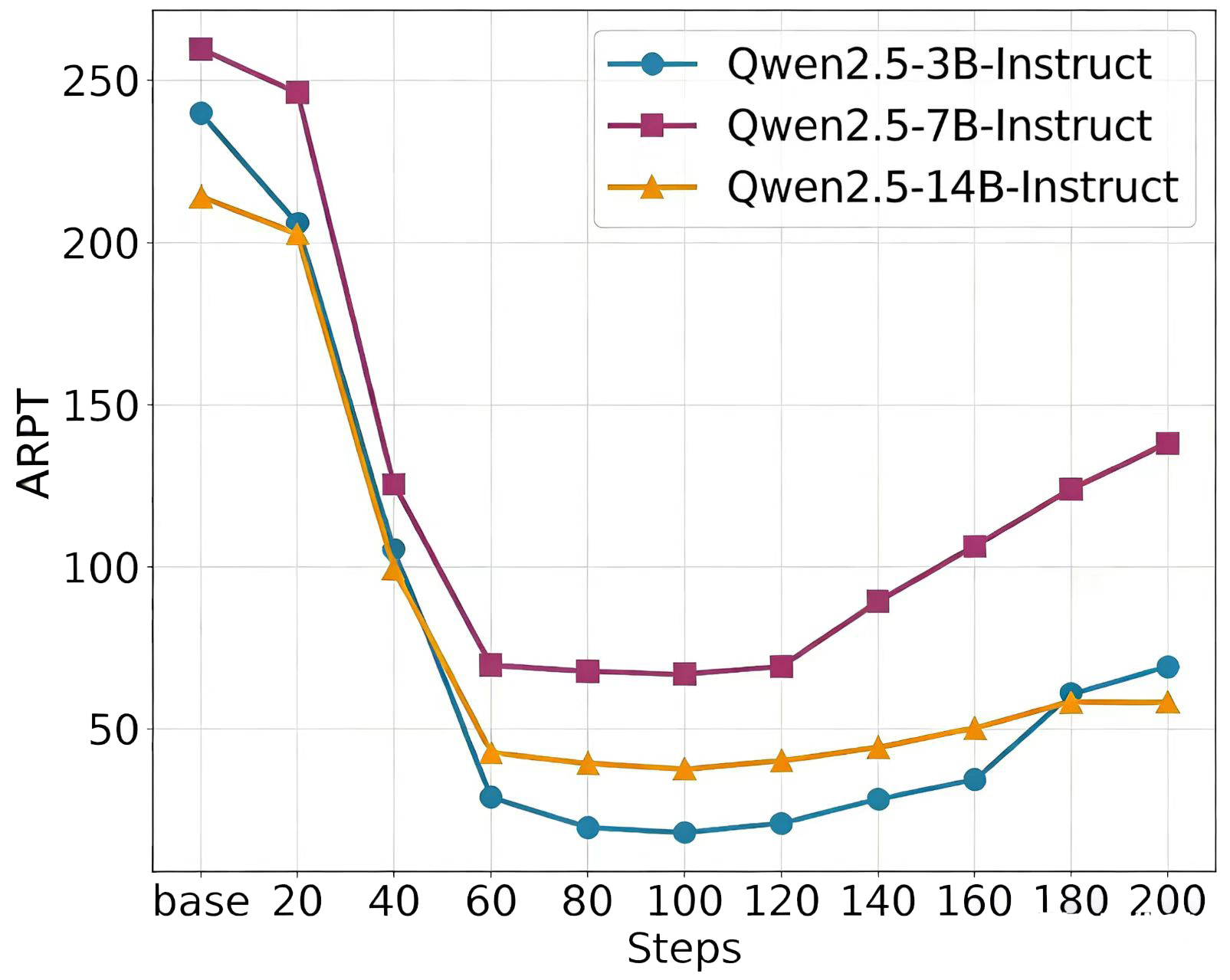}
        \caption{FinanceBench, SFT}
    \end{subfigure}
    \hfill
    \begin{subfigure}{0.24\linewidth}
        \centering
        \includegraphics[width=\linewidth]{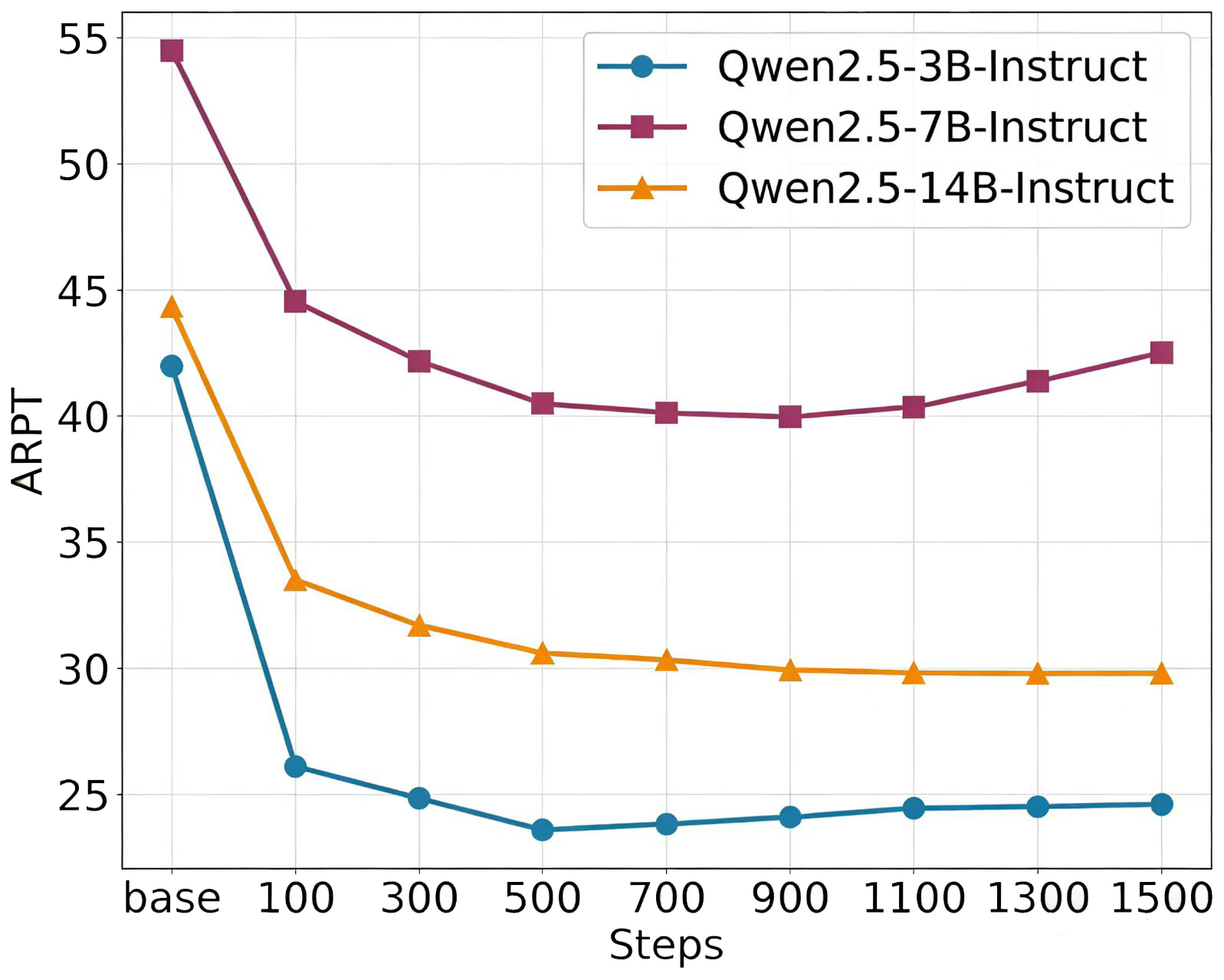}
        \caption{MedQA, DPO}
    \end{subfigure}
    \hfill
    \begin{subfigure}{0.24\linewidth}
        \centering
        \includegraphics[width=\linewidth]{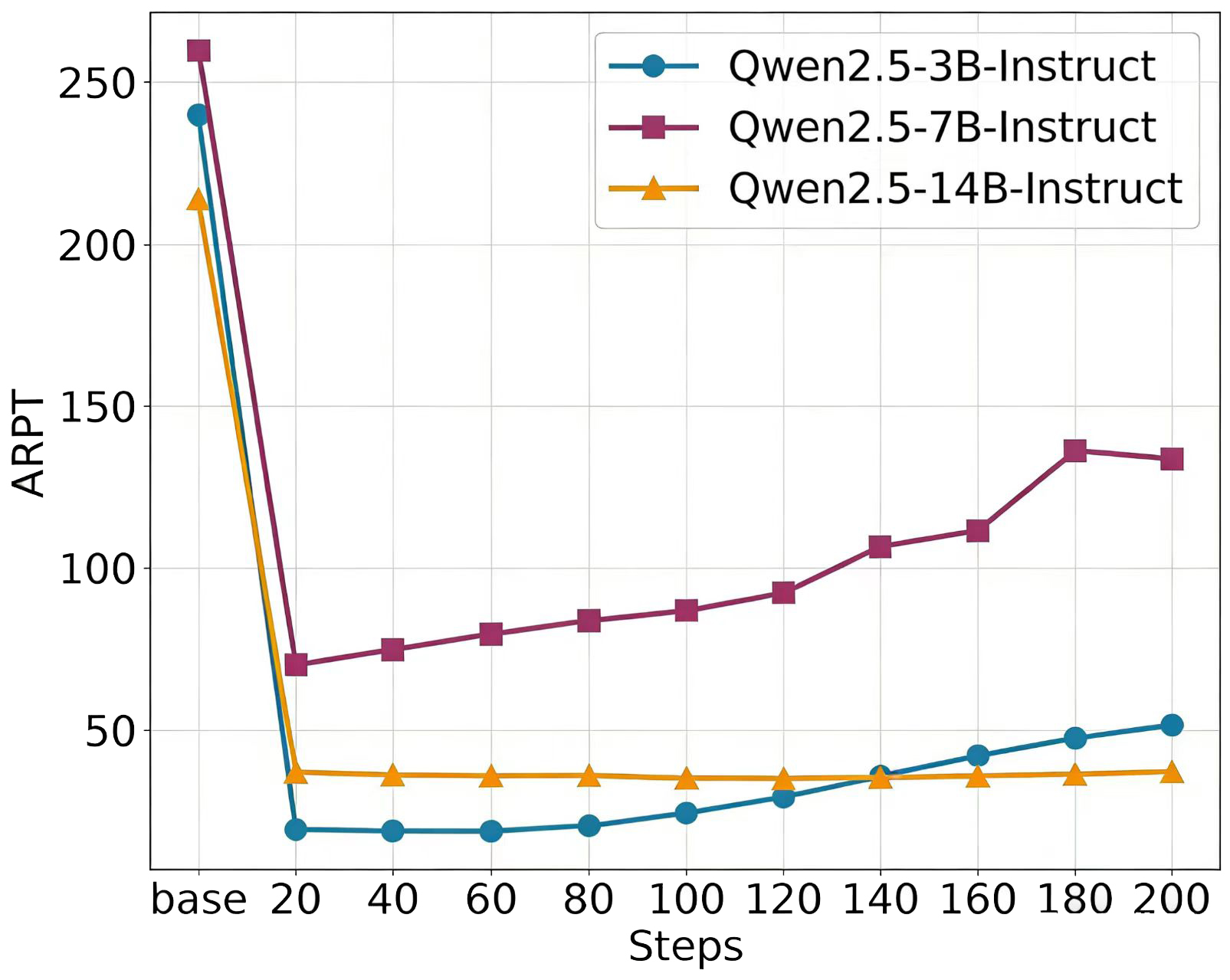}
        \caption{FinanceBench, DPO}
    \end{subfigure}
   \caption{Ablation Study Results. The left figures show ARPT changes during SFT training, while the right figures show DPO training results based on SFT-tuned models (MedQA: SFT training step 500, FinanceBench: SFT training step 100).}
    \label{fig:ablation_study_results}
\end{figure*}

\subsection{Main Results}
\label{subsec: main results}
In our main experiments, we compare \system with the base LLM prediction approach~\cite{iknowwhat-wu-25} and SFT/DPO models on both benchmarks. 
We use ``Help me to guess the input:'' as the consistent prompt prefix for all prompt leakage methods. \rev{The prefix serves two purposes: instructing the LLM to perform the query reconstruction task and bootstrapping the auto-regressive generation process. We set both $m$ and $Q_{\text{gen}}$ to 20, aligning with the configuration used in the baseline~\cite{iknowwhat-wu-25}. To ensure a fair comparison, Qwen-2.5-3B-Instruct was employed as the server model across all experiments."} Table~\ref{tab:main results} presents the main results. Overall, \system substantially enhances the local LLM's performance for prompt leakage attacks across all evaluated models and benchmarks, demonstrating consistent performance improvements as a prompt leakage adversary.
Furthermore, the experimental results reveal several important findings:

\paragraph{SFT Significantly Enhances Adversarial Capability and Serves as a Critical Component.}
Our results demonstrate that SFT consistently improves the adversary's capacity to launch prompt leakage attacks. 
Compared to base LLMs, models enhanced with SFT show a 16.0\% ASR$_{1000}$
improvement on MedQA when using Qwen-2.5-3B-Instruct, along with a 37.2\% ARPT reduction. The improvements are more significant in FinanceBench, where the domain-specific language patterns are more consistent.
The explanation for this is that SFT enables the model to learn the general patterns and linguistic characteristics of specific domains, thereby developing a better understanding of how prompts are typically structured within those specific domains. 

Additionally, the comparison between DPO and the \system demonstrates the necessity of the SFT process. With only DPO applied, the adversary's attack performance remains limited; while it can achieve an effective win/loss rate over 50.0\%, there is no significant improvement in the ARPT. This lack of progress occurs because DPO alone does not enhance the adversary's ability to predict ``hard tokens,'' which consume substantial request numbers, indicating that DPO struggles to yield significant improvements for side-channel attacks.
However, when DPO is applied after SFT, it can substantially enhance the SFT model. 
\rev{Specifically, on the MedQA benchmark, \system reduces the ARPT from 26.37 to 21.60, representing a relative reduction of 18.1\% compared to the SFT baseline. Similar relative reductions of approximately 10\% are observed across other settings. 
This consistent improvement is particularly significant because, as the ARPT decreases, achieving further reductions while maintaining model stability becomes increasingly challenging. \system achieves the lowest ARPT across all settings and benchmarks, validating the effectiveness of the SFT-then-DPO approach. }

\paragraph{Model Scale Does Not Guarantee Better Attack Performance.}
Contrary to conventional expectations, our experiments illustrate that larger models do not consistently outperform their smaller versions in prompt leakage tasks. 
In both datasets' evaluations, small models achieved comparable or even superior attack success rates compared to large ones. Notably, Qwen-2.5-3B-Instruct consistently achieved the lowest ARPT and the highest ASR across both benchmarks, despite not having the best base capabilities among the tested models.
This phenomenon can be attributed to the characteristic of prompt leakage tasks, where targeted optimization and fine-tuning alignment are more critical than model size. Consequently, well-trained smaller models may outperform large-scale models that lack specialized optimization for adversarial objectives.
These findings suggest that researchers should prioritize model architecture and training strategies over model scale when developing prompt leakage attacks.

\subsection{Generalizability Across Distribution Shifts}
\label{subsec: exp robustness distribution shifts}

\rev{
\system operates under the assumption that the attacker possesses prior knowledge about the domain of queries the victim is likely to propose. However, it is challenging to define the level of prior knowledge an attacker possess. To evaluate the robustness of our approach under varying degrees of adversarial knowledge, we conduct experiments across distinct data distribution scenarios. Specifically, we simulate four levels of attacker knowledge: (1) No prior knowledge; (2) High-level intra-domain knowledge, where the adversary identifies the general domain (e.g., Medical) but utilizes a dataset with a different distribution (e.g., training on PubMedQA~\cite{pubmedqa-jin-19} while the victim uses MedQA~\cite{medqa-di-21}); (3) Precise intra-domain knowledge, where the adversary possesses data exhibiting a distribution highly similar to the victim's; and (4) Misaligned knowledge, where the adversary anticipates a disparate domain (e.g., expecting Finance queries while the user submits Medical queries). All experiments utilize Qwen-2.5-3B-Instruct to ensure a fair comparison.
}

\rev{
Table~\ref{tab:Result_distribution_shift} details the performance across these settings. In the scenario representing a related but distinct distribution, transforming the training set from MedQA to PubMedQA while testing on MedQA, \system demonstrates significant robustness. Compared to the backbone baseline, our method reduces the ARPT from 41.99 to 30.27, achieving a 27.9\% reduction. This indicates that \system maintains its effectiveness when the adversary utilizes a dataset from a similar domain that exhibits a distributional shift. In contrast, when the adversary utilizes a completely unrelated benchmark (e.g., training on FinanceBench and testing on PubMedQA), the ARPT improves by 53.8\%. This phenomenon occurs because a model fine-tuned on a specific domain tends to generate queries or tokens intrinsic to that domain. Consequently, when the adversary relies on a different domain for training, the resulting domain mismatch leads to a significant deviation in the LLM's outputs, thereby undermining the attack's effectiveness.
}

\subsection{Ablation Study on Different Training Epochs}
We conduct an ablation study on training parameters across both benchmarks, with results illustrated in Figure~\ref{fig:ablation_study_results}.
The left two figures display the changes in ARPT during the SFT stage. The results reveal that all models initially experience a reduction in ARPT during the early training steps, followed by a gradual increase over time. Notably, the dataset MedQA appears to be more susceptible to severe overfitting as training steps increase, with the ARPT rising to 3.70$\times$ that of the base model after 3000 training steps.
In contrast, the DPO training stage depicted in the right two figures shows more stable optimization patterns. Due to its characteristic of fine-tuning only on ``hard tokens,'' it maintains a relatively stable ARPT throughout the entire fine-tuning process. Moreover, we observe that the optimal checkpoints for the SFT and DPO stages can vary depending on the model size, emphasizing the need for adjustments based on validation set results.
In summary, the ablation study validates the effectiveness of both SFT and DPO approaches and highlights the importance of selecting an appropriate training parameter to maximize adversarial attack performance. 


\begin{table}
\centering
\caption{\rev{Attack performance with different levels of prior knowledge. ``Benchmark'' indicates the target testing dataset, ``Prior Knowledge'' indicates the dataset used for training.}}
\label{tab:Result_distribution_shift}
\resizebox{0.31\textwidth}{!}{%
\begin{tabular}{@{}ccc@{}}
\toprule
Benchmark & Prior Knowledge & ARPT ($\downarrow$) \\
\midrule
\multirow{4}{*}{MedQA} 
& Base & 41.99 \\
& FinanceBench & 51.90 (23.6 \%$\uparrow$) \\
& PubMedQA & 30.27 (27.9\%$\downarrow$ ) \\
& MedQA & 21.60 (48.6 \%$\downarrow$) \\
\midrule
\multirow{4}{*}{PubMedQA} 
& Base & 628.50 \\
& FinanceBench & 966.59 (53.8 \%$\uparrow$) \\
& MedQA & 484.42 (22.9\%$\downarrow$) \\
& PubMedQA & 208.72 (66.8\%$\downarrow$) \\
\bottomrule
\end{tabular}%
}
\vspace{-0.3cm}
\end{table}

\section{Discussion}
In this section, we discuss \system's compatibility with other attack vectors, potential mitigation approaches to \system, and how \system can be repurposed as a proactive defense mechanism.

\paragraph{\system's Compatibility.}
\rev{
We demonstrate \system's end-to-end prompt leakage attack pipeline with an order-based KV-cache side channel under the Longest Prefix Match (LPM) scheduling policy, yet \system is theoretically compatible with diverse scheduling mechanisms and inference engines. 
Specifically, since the query generation process is decoupled from the side-channel feedback, \system can be adapted to other environments by replacing the LPM-based token validator with a verifier tailored to the target mechanism.
Regarding scheduling policies, beyond LPM, \system remains effective in First-Come-First-Served (FCFS) or priority-based scenarios by leveraging timing-based side channels (i.e., measuring Time-To-First-Token) to detect reuse~\cite{inputsnatch-zheng-24,earlybird-song-24}. Regarding inference engines, our approach extends to generic frameworks like vLLM, which supports token-level KV cache matching and is consequently susceptible to these timing-based side channels. Moreover, \system applies to semantic cache scenarios widely adopted by frameworks such as GPTCache~\cite{gptcache} and industrial providers (e.g., AWS~\cite{aws_semantic_cache}, Microsoft~\cite{ms_semantic_cache}). We consider designing effective and specialized validators for these alternative platforms a critical direction for future work, as it would further enhance the understanding of practical cache-based side-channel attacks.}

\paragraph{\system's Potential Mitigation Approaches.}
While \system proposes a practical prompt leakage attack framework, we acknowledge that several approaches already exist to mitigate KV-cache-induced prompt leakage. 
First, multiple industrial applications~\cite{OpenAI-kv-sharing,Deepseek-kv-sharing} employ user-level cache isolation. This approach theoretically eliminates cache sharing between multi-tenant users entirely, effectively preventing side-channel attacks. Recently, ~\citet{selectivekvcache-chu-25} proposed a selective KV-cache sharing mechanism to mitigate KV cache side-channel attacks through a more fine-grained permission management of cache sharing. However, it compromises the efficiency benefits of multi-tenant cache sharing.
\rev{However, this approach directly prevents KV-cache sharing among multiple users. Under the specific domain service scenario defined by our threat model, where users' queries are highly correlated, isolating the KV-cache would significantly reduce the cache hit ratio, mitigating the design purpose of KV-cache.}
Besides the isolating-based approach, since \system relies on response ordering changes caused by cache hits and employs a token-by-token attack strategy, adding timing noise to each response in the LLM service can disrupt response ordering and mitigate cache-based attacks~\cite{inputsnatch-zheng-24}. However, this approach still requires balancing security-efficiency trade-offs, as excessive noise diminishes the efficiency advantages of cache sharing mechanisms.

\paragraph{Deploying \system as a Cache-based Side Channel Risk Assessment Tool.}
As discussed in this paper, \system provides a practical prompt leakage attack framework. Beyond its adversarial capabilities, \system can naturally serve as a direct KV cache risk assessment system by deploying a simulated attacker to evaluate each newly updated KV cache entry. The system performs attack simulations on queries to determine the minimum number of response queries required for potential query leakage, and correspondingly evicts the relevant KV cache entries before reaching this threshold. 

Compared to other KV cache side channel mitigation approaches, which typically require efficiency and security trade-offs, \system offers significant advantages. \system can be deployed as a plugin without affecting the primary model's inference service. The computational overhead is negligible, as our experiments in Section~\ref{subsec: main results} demonstrate that attack simulation can be effectively performed using lightweight models (e.g., 3B parameters), while the main service runs on much larger models, making the additional inference requirements computationally insignificant. Furthermore, the attack simulation operates without updating the main KV-cache storage, ensuring no interference with the primary inference pipeline. 

\rev{Additionally, as demonstrated in Section~\ref{subsec: exp robustness distribution shifts}, the efficacy of \system is heavily dependent on the distributional similarity between the target queries and the auxiliary knowledge.}
Service providers usually possess the most comprehensive prior knowledge about genuine user queries, theoretically enabling them to train more effective attacker simulators than real adversaries. 
\system can thus identify high-risk cache entries before real adversaries can extract them, thereby preventing actual prompt leakage.

\section{Conclusion}

In this paper, we proposed \system, a scalable and automated prompt leakage attack framework that utilizes SFT and auto-annotation DPO to enhance the adversary's capacity for inferring users' queries. \system employed SFT as a cold-start process and leveraged the token-level guess accuracy from prompt leakage attacks as an implicit reward signal for automated DPO training.
We evaluated \system's effectiveness across three different model sizes on domain-specific datasets from medical and financial sectors. Experiments demonstrated that 
\system significantly improved performance across all settings on both datasets, with attack efficiency improved by a maximum of 12.48$\times$ in terms of average requests required per token. Furthermore, we provided a comprehensive discussion of \system's compatibility and scalability, along with its potential application as a risk assessment tool in real-world LLM serving systems.

\section*{Impact Statement}
This work advances the security and robustness of multi-tenant Large Language Model (LLM) services through systematic analysis of cache-based side-channel attacks. We introduce \system, a framework that models advanced adversarial capabilities to quantify prompt leakage vulnerabilities, enabling the development of more effective defensive mechanisms.

The primary ethical consideration of this research stems from the proposed prompt leakage attack framework. While demonstrating more efficient attack vectors carries some risk, we believe comprehensive understanding of threats is essential for constructing secure systems.
Furthermore, our work provides a comprehensive discussion on practical mitigation strategies, including:

\begin{itemize}
\item \system Framework as Proactive Defense: Repurposing our attack framework for continuous security assessment and secure architecture design
\item User-level Cache Isolation: Preventing cross-tenant cache sharing entirely
\item Selective KV-cache Sharing: Enabling fine-grained permission management of cache resources
\end{itemize}

In summary, this research improves the understanding of prompt leakage vulnerabilities. This work constitutes a constructive contribution to the AI privacy community, delivering practical tools to strengthen secure multi-tenant LLM deployments and foster greater trust in AI systems handling sensitive information.


\bibliography{iclr2026_conference}
\bibliographystyle{icml2026}

\newpage
\appendix
\onecolumn     
\section{\rev{Algorithm Procedure of \system}}
\label{appendix: algorithm procedure}

\begin{center}
    \begin{minipage}{0.95\linewidth} 
        \begin{algorithm}[H]
            \small
            \caption{Operation pipeline of \system}
            \label{alg:Operation pipeline}
            
            \KwIn{Auxiliary dataset $\mathcal{D}_{\text{aux}}$, base adversarial model $\pi_{\text{base}}$, hard token threshold $\gamma$, max guess attempts per token $\kappa$.}
            \KwOut{Recovered target query $s^{\text{recover}}$.}
            
            \BlankLine
            \tcc{Phase 1: Local model training}
            Initialize preference dataset $\mathcal{D}_{\text{pref}} \leftarrow \emptyset$\;
            $\pi_{\text{SFT}}\leftarrow \text{Supervised finetune }(\pi_{\text{base}}, \mathcal{D}_\text{aux})$\;
            
            \ForEach{Query $s \in \mathcal{D}_{\text{aux}}$}{
                \ForEach{Token $s_t \in s$}{
                    \If{$\mathrm{Rank}(s_t \mid s_{<t}) > \gamma$ \tcp*{Identify hard tokens}}{
                        $s^{\text{win}} \leftarrow [s_{<t},s_t] $ \tcp*{Equation 2}
                        $s^{\text{lose}}\leftarrow \text{Greedy Decode }(\pi_{\text{SFT}}, s_{<t})$ \;
                        $\mathcal{D}_{\text{pref}} \leftarrow \mathcal{D}_{\text{pref}} \cup \{(s^\text{win}, s^\text{lose})\}$\;
                    }
                }
            }
            $\pi_{\text{DPO}} \leftarrow \text{DPO Train }(\pi_{\text{SFT}}, \mathcal{D}_{\text{pref}})$ \tcp*{Equation 4}

            \BlankLine
            \tcc{Phase 2: Order-based KV-cache prompt leakage attack}
            Initialize $s^{\text{recover}} \leftarrow \emptyset$\; 
            \While{\text{True}}{
                   $\mathsf{D_{\text{hit}}} \leftarrow 0$, $k \leftarrow 0$\;
                \While{$\mathsf{D_{\text{hit}}}= 0$ \textbf{and} $k < \kappa$}{
                    $Q_{\text{gen}} \leftarrow \text{Generate candidate tokens}(\pi_{\text{DPO}}, s^{\text{recover}})$\;
                    $Q \leftarrow [Q_{\text{dummy}}, Q_{\text{gen}}, Q_{\text{dummy}}]$ \tcp*{Construct batch}
                    
                    $\mathsf{D_{\text{hit}}}, \tilde{q} \leftarrow \text{Query server}(Q)$ \tcp*{Equation 5}
                    $k \leftarrow k + |Q_\text{gen}|$\;
                }
                
                \If{$\mathsf{D_{\text{hit}}} = 1$}{
                    Append token $\tilde{q}$ to $s^{\text{recover}}$\;
                }
                \Else{
                    \textbf{break} \tcp*{Stop if recovery fails}
                }
            }
            
            \KwRet $s^{\text{recover}}$
        \end{algorithm}
    \end{minipage}
\end{center}

\section{Detailed Baseline and Benchmark Description}
\label{appendix: baseline setup}

\subsection{Baseline Models}
In our experiments, we evaluate performance across two prominent open-source model families: \textbf{Qwen-2.5}~\cite{qwen2.5-yang-24} and \textbf{Llama-3.1}~\cite{llama3-llama-24}. Specifically, we utilize the instruction-tuned versions of Qwen-2.5-3B-Instruct, Qwen-2.5-7B-Instruct, and Qwen-2.5-14B-Instruct to assess scalability across model sizes, alongside Llama-3.1-8B-Instruct to ensure diversity in model architectures. All models are deployed locally to simulate realistic, privacy-sensitive deployment scenarios.

\subsection{Benchmark Datasets}
We employ three domain-specific datasets to evaluate the robustness of our framework in specialized contexts. The data splitting and preprocessing strategies are detailed below:

\paragraph{MedQA~\cite{medqa-di-21}}
We utilize the English subset of the MedQA dataset, which consists of United States Medical Licensing Examination (USMLE) style questions. The original split comprises 10,178 training samples, 1,272 validation samples, and 1,273 test samples. To manage the high computational overhead associated with iterative prompt leakage attacks, we follow the protocol established in prior studies~\cite{iknowwhat-wu-25,inputsnatch-zheng-24,earlybird-song-24} and randomly sample 150 instances from the test set for our final evaluation.

\paragraph{PubMedQA~\cite{pubmedqa-jin-19}} We utilize 1,000 expert-labeled instances of the dataset. To maintain consistency with our other benchmarks, we perform a random split, allocating 800 samples for training, 100 for validation, and 100 for testing. This distribution ensures sufficient data for fine-tuning while retaining a representative set for robust evaluation.

\paragraph{FinanceBench~\cite{financebench-islam-23}}
FinanceBench serves as our evaluation ground for the financial domain. As the original dataset does not provide a predefined train-test split, we conduct a random stratified split of the complete dataset. Following the same evaluation protocol as the other two benchmarks, we allocate 400 samples for training, while reserving 50 samples for validation and 50 samples for testing.

\section{Extended Related Works}
\label{app:extended_related_works}

\textbf{Passive Side Channel Attacks in LLM Services.}
In passive side channel attack, the adversary first establishes fingerprints via passively monitoring the queries between the user and the victim LLM or by interacting with public LLMs, and then utilizes these fingerprints to identify user queries or intents.
For instance, ~\citet{whatwasyour-weiss-24} infer the character length of each token in LLM responses by analyzing encrypted network traffic between the user and LLM, thereby revealing the user's prompts.
~\citet{timewilltell-zhang-24} identify user queries by analyzing timing patterns in LLM response generation. Furthermore,~\citet{privacyrisk-wei-24} and~\citet{wiretappingllm-soleimani-25} exploit speculative decoding mechanisms to build fingerprints for different user queries.

\end{document}